\documentclass[twocolumn,prl,letterpaper,showpacs]{revtex4}
\usepackage{amsmath}
\usepackage{graphicx}
\usepackage{amssymb}

\newcommand{\pq}[2]{#1\,{\rm #2}}
\renewcommand{\Re}{\operatorname{Re}}
\renewcommand{\Im}{\operatorname{Im}}
\newcommand{\Ej}{E_{\rm J}}
\newcommand{\Tn}{T_{\rm N}}
\newcommand{\Rn}{R_{\rm N}}

\newcommand{\kB}{k_{\rm B}}
\newcommand{\Hj}{\hat H_{\rm J}}
\newcommand{\pairrate}{\Gamma^{\rm Cp}}\newcommand{\photonrate}{\Gamma^{\rm ph}}
\newcommand{\znorm}{r}
\newcommand{\density}{\gamma}
\newcommand{\prefactor}{\rho}
\newcommand{\occup}{p}
\newcommand{\eV}{\mathnormal{e}V}
\newcommand\nuJ{\nu_{\rm J}}

\newcommand\ket[1]{| #1 \rangle}
\newcommand\bracket[3]{\left\langle #1 \vphantom{{#2}{#3}}\right| #2 \left| #3 \vphantom{{#1}{#2}}\right\rangle}

\begin{document}

\title{The Bright Side of Coulomb Blockade}

\author{M. Hofheinz}

\author{F. Portier}

\author{Q. Baudouin}

\author{P. Joyez}

\author{D. Vion}

\author{P. Bertet}

\author{P. Roche}

\author{D. Esteve}

\affiliation{Service de Physique de l'Etat Condens\'e (CNRS URA 2464), IRAMIS, CEA Saclay, 91191 Gif-sur-Yvette, France}

\date{\today}
\begin{abstract}
We explore the photonic (bright) side of dynamical Coulomb blockade (DCB) by measuring the radiation emitted by a dc voltage-biased Josephson junction embedded in a microwave resonator. In this regime Cooper pair tunneling is inelastic and associated to the transfer of an energy $2eV$ into the resonator modes. We have measured simultaneously the Cooper pair current and the photon emission rate at the resonance frequency of the resonator. Our results show two regimes, in which each tunneling Cooper pair emits either one or two photons into the resonator. The spectral properties of the emitted radiation are accounted for by an extension to DCB theory.

\end{abstract}
\maketitle 

\begin{figure}
\includegraphics[width=0.95\columnwidth]{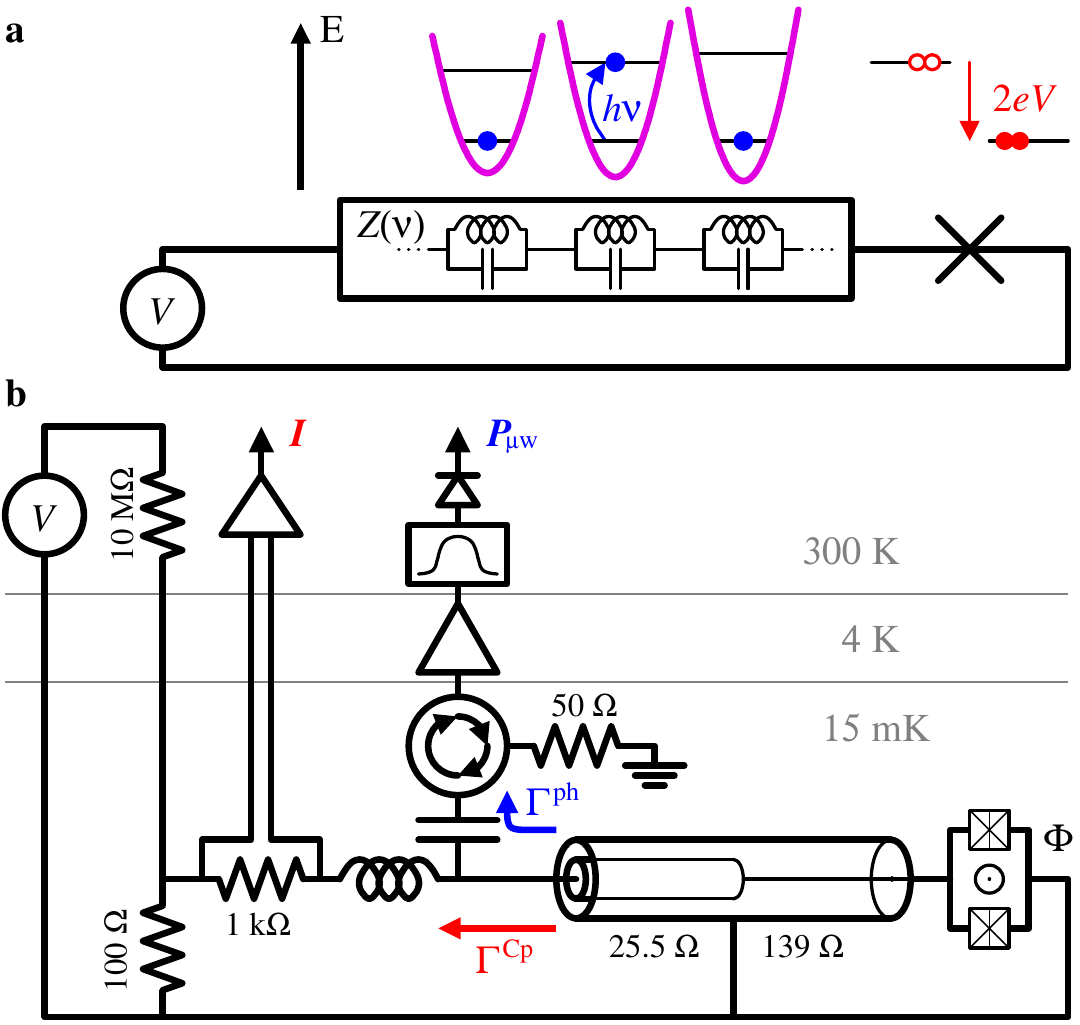} \caption{\label{fig:scheme} Dynamical Coulomb blockade of Josephson tunneling. {\bf a}--Generic circuit: a dc voltage-biased Josephson element in series with an electromagnetic environment with impedance $Z(\nu)$, which can be described as a collection of harmonic oscillators. Below the gap voltage, the energy $2eV$ of a tunneling Cooper pair is transferred to the electromagnetic environment in form of one or several photons in its  modes. {\bf b}--Experimental setup: The sample consists of a SQUID, working as a tunable Josephson junction, in series with a microwave resonator, forming the electromagnetic environment and consisting of two quarter-wave transformers. A bias tee separates the low frequency voltage bias $V$ and current measurement $I$ from the emitted microwave power $P_{\rm \mu w}$. Three circulators (only one represented here) ensure thermalization of the microwave environment. The emitted power is amplified at 4.2~K and then bandpass filtered and detected at room temperature.}
\end{figure}

Dynamical Coulomb blockade (DCB) of tunneling is a quantum phenomenon in which tunneling of charge through a small tunnel junction is modified by its electromagnetic environment \cite{devoret90,girvin90,averin90,ingold92}. This environment is described as an impedance in series with the tunnel element (see Fig.~\ref{fig:scheme}a). The sudden charge transfer associated with tunneling can generate photons in the electromagnetic modes of the environment. In a normal metal tunnel junction, biased at voltage $V$, the energy $eV$ of a tunneling electron can be dissipated both into quasiparticle excitations in the electrodes and into photons.  At low temperature energy conservation forbids tunneling processes emitting photons with total energy higher than $eV$. This suppression reduces the conductance at low bias voltage \cite{devoret90,girvin90,ingold92}. In a Josephson junction, DCB effects are more prominent since at bias voltages smaller than the gap voltage $2\Delta/e$ quasiparticle excitations cannot take away energy. Therefore, as shown in Fig.~\ref{fig:scheme}a, the entire energy $2eV$ of tunneling Cooper pairs has to be transformed into photons in the impedance for a dc current to flow through the junction \cite{averin90,ingold92}. Experiments have confirmed the predictions of DCB theory for the tunneling current, both in the normal \cite{delsing89, geerligs89,cleland90} and superconducting case \cite{holst94,basset10} but the associated emission of photons into the environment has never been investigated. The aim of this work is precisely to fill this gap by exploring the photonic side of DCB. We do so by embedding a Josephson junction into a well controlled electromagnetic environment provided by a microwave resonator. The resonator in turn leaks photons into an amplifier, allowing to measure the rate and spectrum of photons emitted by the junction. 

\begin{figure}
\includegraphics[width=\columnwidth]{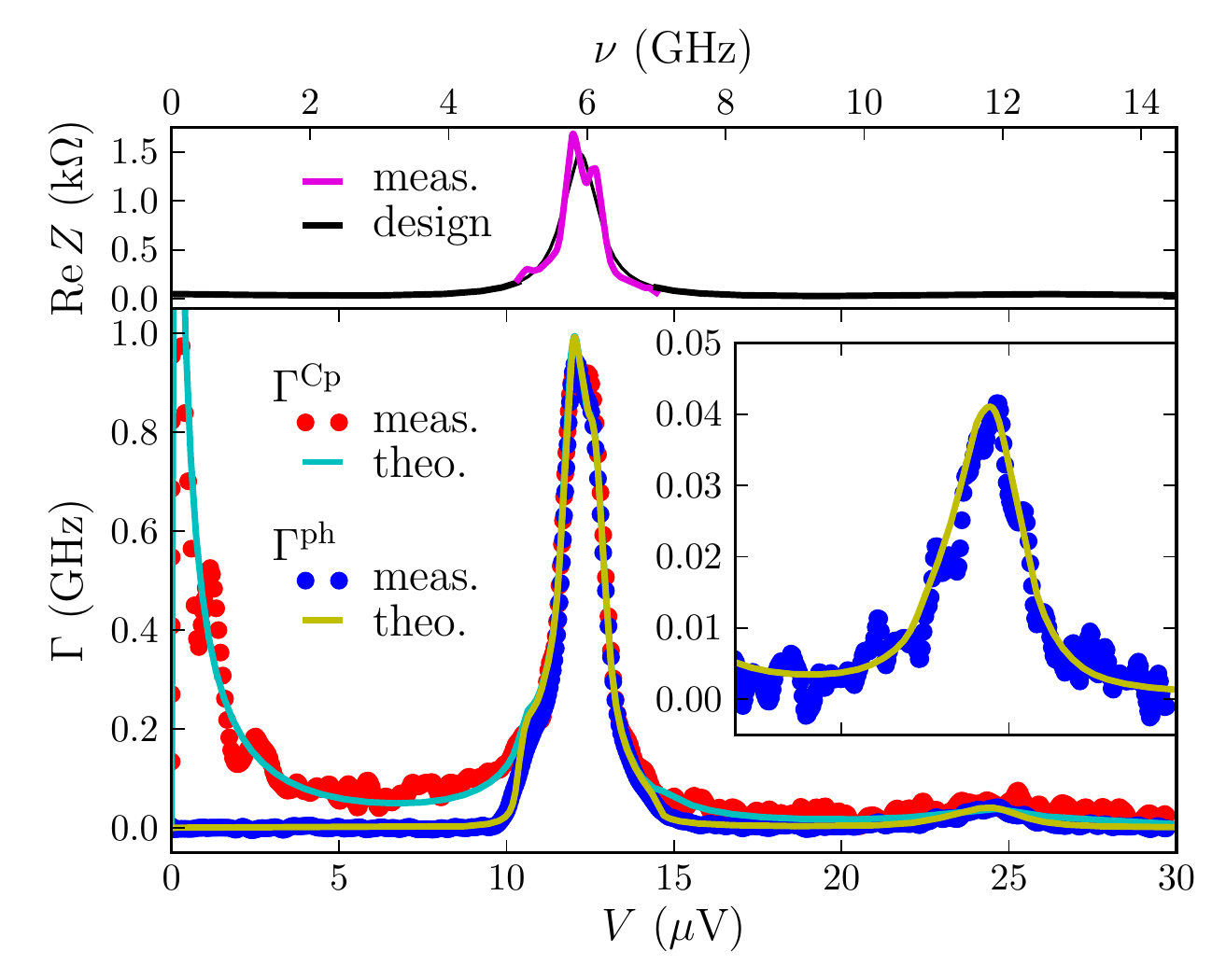} 
\caption{\label{fig:IP} Fundamental mode $\nu_0$ of the resonator and corresponding Cooper pair and photon rates. Top panel: real part of the impedance seen by the junction, calculated from the resonator geometry (black line) and  reconstructed (magenta) from a quasiparticle shot noise measurement \cite{si}. Bottom panel: measured Cooper pair rate $\pairrate$ (red) and photon rate $\photonrate_{0}$ (blue) integrated from 5 to $\pq{7}{GHz}$. Both rates show a peak around $V = h\nu_0/2e \simeq \pq{12}{\mu V}$. A second small peak in the microwave power at $V=\pq{24}{\mu V}$ (vertical zoom in inset) corresponds to two-photon processes. Solid lines are
$P(E)$ theory fits using the reconstructed impedance from 5 to $\pq{7}{GHz}$ and the calculated one outside this range, an effective temperature of $\pq{60}{mK}$, and a single adjustable parameter $\Ej = \pq{5.1}{\mu \eV}$. Cyan and yellow lines  correspond, respectively, to Eq.~(\ref{eq:pairrate}) and Eq.~(\ref{eq:photonrate}) integrated from 5 to $\pq{7}{GHz}$.}
\end{figure}

\begin{figure}
\includegraphics[width=1\columnwidth]{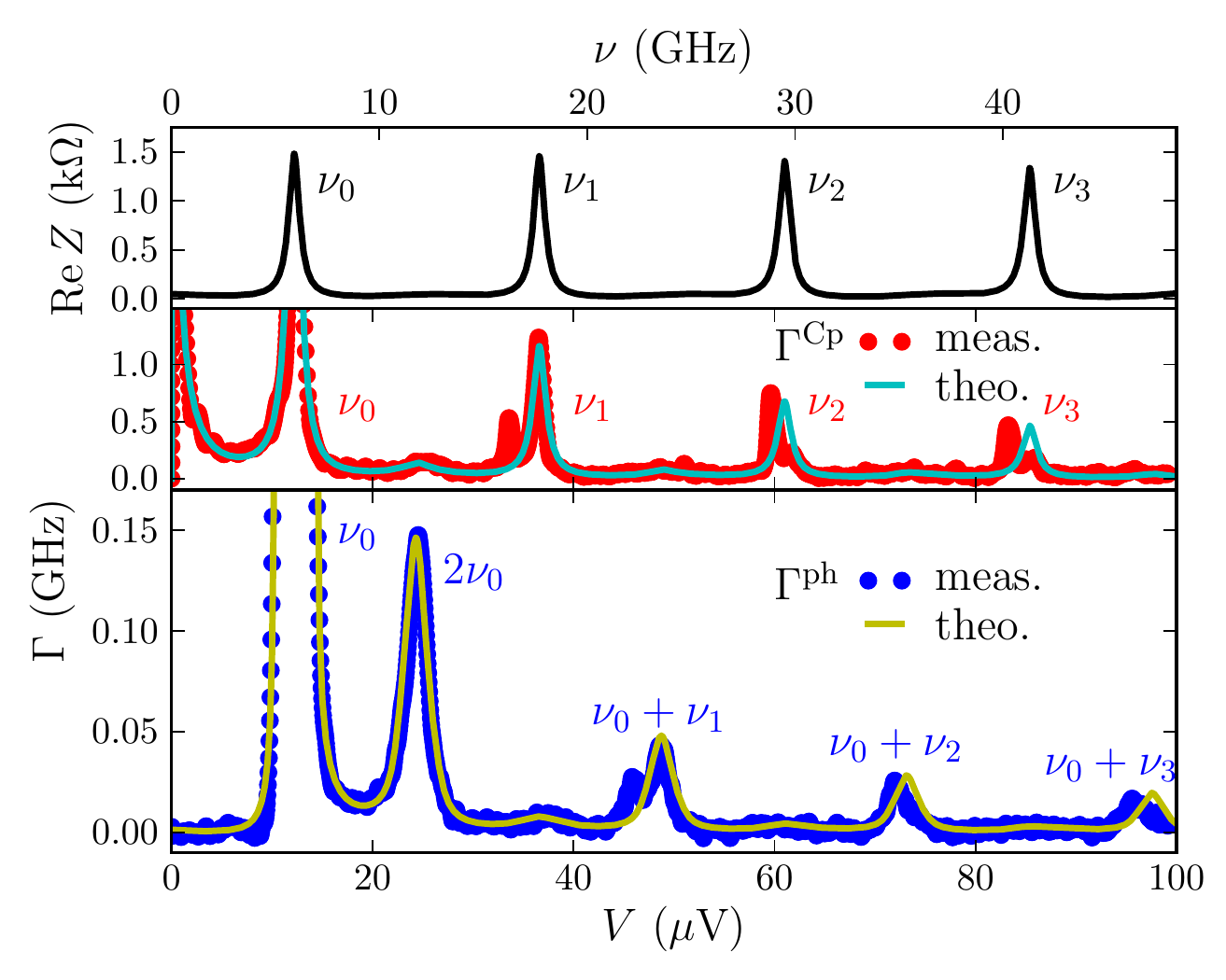} \caption{\label{fig:multi} Higher orders modes $\nu_n$ of the resonator and corresponding Cooper pair and photon rates. The plotted quantities are the same as in Fig.~\ref{fig:IP}, but for a larger voltage and frequency range and different Josephson coupling. The Cooper pair rate shows peaks at voltages $V=h\nu_n/2e$, associated to the emission of photons in the resonator mode at frequency $\nu_n$. The power measurement only probes the photons emitted in the fundamental mode of the resonator. It shows a large peak at $V=h\nu_0/2e$, and smaller ones at $V=h(\nu_0+\nu_n)/2e$, associated to the simultaneous emission of one photon at $\nu_0$ and one photon at $\nu_n$ per tunneling Cooper pair. Here theoretical curves are obtained from the calculated impedance in the whole frequency range and the fitting parameter $\Ej = \pq{10.3}{\mu \eV}$.} 
\end{figure}

The experimental setup is represented in Fig.~\ref{fig:scheme}b. A small SQUID acts as a tunable Josephson junction with Josephson energy $\Ej=E_{{\rm J} 0}|\cos(e\Phi/\hbar)|$ adjustable via the magnetic flux $\Phi$ threading its loop. The microwave resonator is made of two quarter-wave transformers and its fundamental mode has frequency $\nu_{0}\simeq\pq{6.0}{GHz}$ and quality factor $Q_{0}\simeq9.4$. Higher modes of the resonator appear at $\nu_{n}\simeq(2n+1)\nu_{0}$ ($n=1,2,\ldots$) with the same lineshape up to small deviations caused by the junction capacitance (estimated to be $\pq{4}{fF}$). The expected total impedance $Z(\nu)$ seen by the Josephson element is plotted in the top panels of Figs.~\ref{fig:IP} and \ref{fig:multi}, and reaches $\simeq \pq{1.5}{k\Omega}$ for all modes.

The sample is cooled to $T\simeq\pq{15}{mK}$ in a dilution refrigerator and connected to a bias tee separating the high-frequency
and low-frequency components of the current. The low-frequency port
is connected to a voltage bias through a $\pq{1}{k\Omega}$ resistor
used to measure the tunneling current $I$. The microwave port is connected
to a 4 to $\pq{8}{GHz}$ cryogenic amplifier with noise temperature $\Tn\simeq\pq{3.5}{K}$ through three circulators protecting the sample from the amplifier noise and ensuring thermalization of the environment. 
After further amplification and eventually bandpass filtering at room temperature, 
the power is detected using a calibrated square-law detector. Its output voltage is proportional
to the microwave power, which contains the weak sample contribution $P_{\rm \mu w}$ on top of the large noise floor of the cryogenic amplifier. In order to measure $P_{\rm \mu w}$, we remove the background with a low frequency lock-in technique by applying a $0$ to $V$ square-wave bias voltage
modulation to the sample.

\begin{figure}
\includegraphics[width=1\columnwidth]{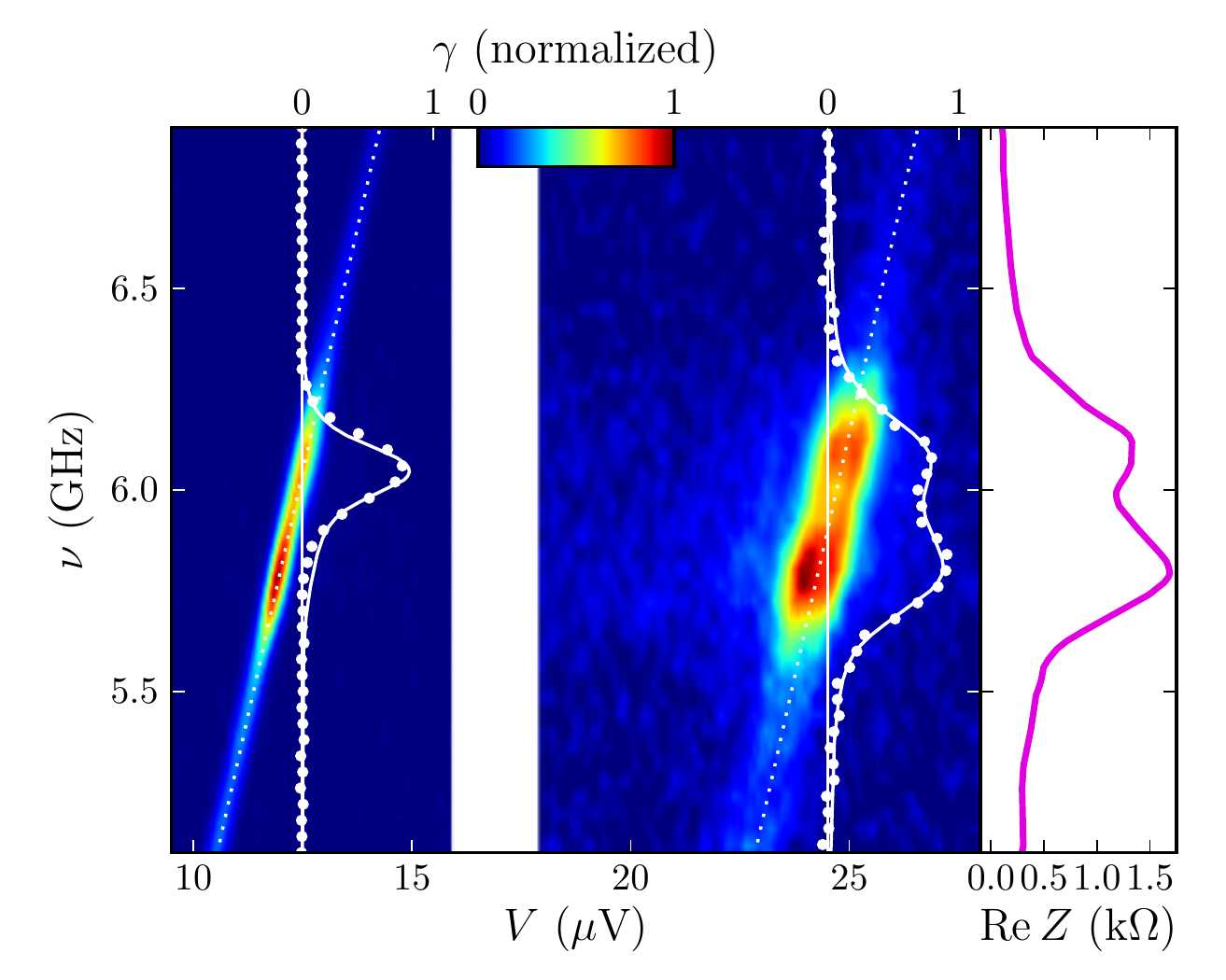} \caption{\label{fig:spectro} Main panel: spectral density of the photon emission rate $\density$ for the first and second order peaks ($\nu_0$ and $2\nu_0$) as a function of bias voltage and frequency. Data were taken for different Josephson energies and were normalized by the maximum $\density$ of each data set.
Dotted lines correspond to $2eV=h\nu$ and $2eV=h(\nu+\nu_0)$. White dots correspond to power spectra at fixed voltages (12.5 and $\pq{24.5}{\mu\eV}$) and solid lines to theoretical predictions  Eq.~(\ref{eq:photonrate}) integrated over $\pq{100}{MHz}$, the width of the filter used in this experiment with an effective temperature of $\pq{60}{mK}$ and fitting parameters $\Ej = \pq{5.0}{\mu \eV}$ at $V=\pq{12.5}{\mu V}$ and $\Ej = \pq{14.3}{\mu \eV}$ at $V=\pq{24.5}{\mu V}$. Right panel: reconstructed dissipative part of the resonator impedance, as in Fig.~\ref{fig:IP}.}
\end{figure}

We first characterize the on-chip microwave resonator and determine
the gain of the microwave chain by measuring the power emitted
by the electronic shot noise of the junction $S_{II}\simeq eI$ at bias voltage $V\simeq\pq{2}{mV}$,
well above the gap voltage $2\Delta/e \simeq \pq{0.4}{mV}$. Under these conditions, 
the spectral density of the emitted power 
is $2eV\!  \Re Z(\nu)\Rn/ |\Rn+Z(\nu)|^{2}\simeq 2eV \Re Z(\nu)/\Rn$
 with $\Rn=\pq{17.9}{k\Omega}\gg|Z(\nu)|$ the tunnel resistance of the junction in the normal state. 
This spectral density is extracted using a heterodyne measurement equivalent to a $\pq{100}{MHz}$ wide bandpass at tunable frequency.
More details about this procedure can be found in \cite{si}. The extracted $\Re Z(\nu)$ 
is shown in the upper panel of Fig.~\ref{fig:IP} and the right panel of Fig.~\ref{fig:spectro}. 
It shows the expected peak at $\nu_0$ with an additional modulation due to parasitic reflections in the microwave set-up.

Once the system calibrated, we measure the Cooper pair transfer rate across the SQUID $\pairrate = I/2e$ and the photon emission rate into the fundamental mode of the resonator $\photonrate_0$ as a function of bias voltage. Rate $\photonrate_0$ is extracted from the microwave power $P_{\rm \mu w}$, emitted 
in a $\pq{2}{GHz}$ wide band centered at $\nu_0 = \pq{6}{GHz}$, via $\photonrate_0=P_{\rm \mu w}/h\nu_0$. In order to simplify interpretation of our data, we adjust $\Ej$ for each experiment to ensure good signal to noise ratio while keeping the maximum $\photonrate_0$ smaller than the decay rate of the resonator $2\pi \nu_0/Q_0\simeq \pq{4}{GHz}$, so that the electromagnetic environment of the junction stays close to thermal equilibrium.  

The measured rates $\pairrate$ and $\photonrate_0$ are shown in Figs.~\ref{fig:IP} and \ref{fig:multi}. In Fig.~\ref{fig:IP} both rates display a peak at voltage $V_0 = h\nu_0/2e=\pq{12}{\mu V}$, where the energy of a tunneling Cooper pair corresponds to the energy of one photon at frequency $\nu_0$. On resonance $\pairrate$ and $\photonrate_0$ agree within $\pq{5}{\%}$. In Fig.~\ref{fig:multi}  additional peaks appear in $\pairrate$ at voltages $V_n=h\nu_n/2e$ $(n=1,2,3)$, which do not have counterparts
in $\photonrate_0$. In both figures $\pairrate$ and $\photonrate_0$ show much smaller peaks at $V_0 + V_n$ $(n=0,1,2,3)$.

DCB theory accounts for these results. At low temperature $k_{B}T\ll2eV$, when backward tunneling
is negligible, the Cooper pair rate is given by \cite{averin90,ingold92}:
\begin{equation}
 \pairrate=\frac{\pi}{2\hbar}E_{J}^{2}P(2eV),
 \label{eq:pairrate}
\end{equation}
where $P(E)$ is the probability density for a tunneling charge $2e$ to emit an energy $E$ in form of photons into the impedance. The function $P(E)$ is the Fourier transform of $\exp J(t)$ with \cite{ingold92}
\begin{equation}
J(t) = 2 \intop_{-\infty}^\infty d\nu \frac{\znorm(\nu)}{\nu} \frac{\exp(-i2\pi\nu t) -1}{1-\exp(-h\nu/\kB T)},
\end{equation}
where
\begin{equation}
\znorm(\nu)=\frac{4e^{2}}{h}\Re Z(\nu).
\end{equation}
At $T=0$, the expansion of the function
$P(E)$ in powers of $\znorm(\nu)$ yields: 
\begin{equation}
\pairrate  \simeq 
\frac{{\Ej^*}^2}{2\hbar^2}\left(\frac{\znorm(\nuJ)}{\nuJ}+\int_{0}^{\nuJ}\!\! d\nu\frac{\znorm(\nu)}{\nu}\frac{\znorm(\nuJ-\nu)}{\nuJ-\nu}\right)\!,\label{eq:pairrateO2}
 \end{equation}
 where $\Ej^*=\Ej\left(1-\int\!d\nu\,\znorm(\nu)/\nu\right)$ and $\nuJ = 2eV/h$. We neglect here the low-frequency environment and suppose $\int\!d\nu\,\znorm(\nu)/\nu \ll 1$. The first term in the parenthesis of Eq.~(\ref{eq:pairrateO2}) corresponds to the emission of a single photon at frequency $\nuJ$. It accounts for the peaks in $\pairrate$ observed at bias voltages $V_n$, and for the peak in $\photonrate_0$ at $V_0$. Since we only detect photons emitted in the fundamental mode, $\photonrate_0$ does not show any peak at $V_n$ for $n>0$. The second term in the parenthesis corresponds to the simultaneous emission of two photons at frequencies $\nu$ and $\nuJ-\nu$. It accounts for the peaks in $\photonrate_0$ observed at bias voltage $V_n + V_0$, where the energy of a tunneling Cooper pair can be split into one photon at frequency $\nu_0$ and another at frequency $\nu_n$. These second order processes have much lower rates than the first order peaks in $\pairrate$ because of the low resonator impedance. The same processes also appear in $\pairrate$ as small peaks between the main resonances. Note however that first order processes contribute about equally to the peaks in $\pairrate$ due to small rises in the impedance at $\nu_0 + \nu_n$. Eq.~\ref{eq:pairrateO2} also accounts for the increase in $\pairrate$ at low voltage in Fig.~\ref{fig:IP} and the reduction of the height of the resonances at $V_n$ with increasing $n$, which both reflect the $1/\nu$ scaling of the first order term. Note that as far as single photon processes are concerned, the ac Josephson effect predicts the same voltage to frequency conversion and Cooper pair and photon rates. However, DCB assumes that the transferred charge is almost a classical variable and its conjugate variable, the phase across the junction, has large quantum fluctuations, whereas the opposite assumption is made in the ac Josephson effect.

In Fig.~\ref{fig:spectro} we explore the spectral properties of the emitted radiation, using a heterodyne technique with $\pq{100}{MHz}$ resolution. The first order emission occurs along a narrow line at $2eV=h\nu$ whose amplitude is modulated by $\znorm(\nu)$, in agreement with Eq.~(\ref{eq:pairrateO2}). This equation, however, makes no prediction about the spectral width of the emitted radiation which is found to be approximately $\pq{150}{MHz}$, significantly narrower than the $\pq{700}{MHz}$ wide resonance in $\znorm(\nu)$. The second order peak follows a $2eV=h(\nu+\nu_{0})$ line and has a much broader spectral width, comparable to the width of the resonance in $\znorm(\nu)$.

In order to account for these observations we calculate the spectral density of photon emission $\density(\nu)$ in the DCB regime, which is related to the  emission current noise spectral density $S_{II}$ via $\density=2\Re Z(\nu) S_{II}(\nu)/h\nu$. From the relation between $S_{II}$ and $P(E)$ \cite{averin90,si} one obtains
\begin{equation}  
\density= \frac{2\znorm(\nu)}{\nu}\cdot \frac{\pi}{2\hbar}\Ej^{2}P(2eV-h\nu).\label{eq:photonrate}
 \end{equation}
This expression, valid at low temperature $k_B T \ll h\nu, 2eV$, can be understood within a Caldeira-Leggett decomposition \cite{caldeira83} of the impedance $Z(\nu)$ into modes with infinitesimal width \cite{si}:  The term $2\znorm(\nu)/\nu$ describes the probability density for each tunneling Cooper pair to emit a photon at frequency $\nu$ into the environment. The remainder of Eq.~(\ref{eq:photonrate}) gives the Cooper pair rate while emitting the remaining energy $2eV-h\nu$ in the form of one or several other photons, as in Eq.~(\ref{eq:pairrate}).

The first order peak occurs at $2eV\simeq h\nu$, so that $P(E)$ has to be evaluated around $E\simeq0$ where the perturbative expression (\ref{eq:pairrateO2}) is not valid. Therefore we calculate $P(E)$ at all orders in $\znorm$ at finite temperature, following the method described in Ref.~\cite{ingold91}. $P(E\simeq 0)$ is approximately a Lorentzian $\frac{D/\pi}{D^{2}+E^{2}}$ of width $2 D\simeq4\pi kT\znorm(0)$ \cite{ingold91}. This expression gives the spectral width of the first order peak. Its height along the line $2eV=h\nu$ follows the variations of $\znorm(\nu)$ due to the first term in Eq.~(\ref{eq:photonrate}). 
In order to find a quantitative agreement with the data we use an effective temperature $T=\pq{60}{mK}$, in good agreement with the $\pq{60}{mK}\pm\pq{10}{mK}$ electronic
temperature deduced from shot noise measurements
performed at $\nu_0$ on the same junction, driven to the
normal state by applying a $\pq{0.1}{T}$ magnetic field \cite{Schoelkopf97}. We attribute the
difference between the effective temperature of the environment and
the fridge temperature of $\pq{15}{mK}$ mainly to low frequency noise radiated by the amplifier used for the current measurement. 

For the second order peak $P(E)$ has to be evaluated around $E\simeq h\nu_{0}$,
where it reproduces the variations of $\znorm(E/h)$. The
radiation of the second order peak is therefore emitted in a band
around $2eV=h(\nu_{0}+\nu)$ with a spectral width given by the width of the resonance in $\znorm(\nu)$. As for the first-order peak, the amplitude along this line is then modulated by $\znorm(\nu)$ due to the first term in Eq.~(\ref{eq:photonrate}). 

With the spectral distribution $\density$ at hand we now come back to the data presented in Figs.~\ref{fig:IP} and \ref{fig:multi} for a more quantitative description. The solid yellow lines in both figures show expression~(\ref{eq:photonrate}), integrated from 5 to $\pq{7}{GHz}$, and the solid cyan lines  expression~(\ref{eq:pairrate}) \endnote{The theory line corresponds in fact to 
$\frac{\pi}{2\hbar}E_{J}^{2}\left(P(2eV)-P(-2eV)\right)$, 
but the additional backward tunneling term is only important below approximately $\pq{1}{\mu V}$.}. The theory accurately matches the measurement with a single fitting parameter, the Josephson energy $\Ej$, and accounts for the relative weight of the first and second order peaks in Fig.~\ref{fig:IP}. Discrepancies only show up at the larger bias voltages in Fig.~\ref{fig:multi}, corresponding to frequencies above $\pq{18}{GHz}$ where we do not accurately
control the electromagnetic environment.
For the sake of consistency with Fig.~\ref{fig:spectro} we have used the non perturbative $P(E)$, although the zero temperature perturbative expression (\ref{eq:pairrateO2}) would already have given a satisfactory agreement with the data. 

In conclusion, our results clearly validate the understanding of DCB based on photon emission into the electromagnetic environment. We have measured the spectral density of the emitted radiation, which we quantitatively account for by an extension to DCB theory. A natural continuation of this work is the investigation of tunneling current and emitted radiation at higher Josephson energy, where the emission rate can be strong enough to overcome cavity loss and cause stimulated emission. A larger Josephson energy will also make  simultaneous tunneling of multiple Cooper pairs relevant. These are the two ingredients for the transition from the DCB regime to the regime of the ac Josephson effect.
Furthermore, our results demonstrate that under appropriate biasing, the tunneling of a Cooper pair is associated with the simultaneous emission of multiple photons. This opens the way to new schemes for the emerging field of microwave quantum optics using inelastic tunneling of Cooper pairs for the generation of non classical photon states. 

We gratefully acknowledge support from the European project Scope, the ANR project Masquelspec, C'Nano IdF, technical help from P. Senat, P. Jacques, P. Orfila and discussions within the Quantronics group.

\clearpage

\section{Supplementary Material}

\subsection{Details on setup and calibration}
We describe the calibration of the low frequency circuitry for voltage
bias and current measurement and the microwave components used to
define the environment of the junction and to measure the emitted
radiation.

\subsubsection{Low frequency circuit}
\begin{figure}
\includegraphics[width=\columnwidth]{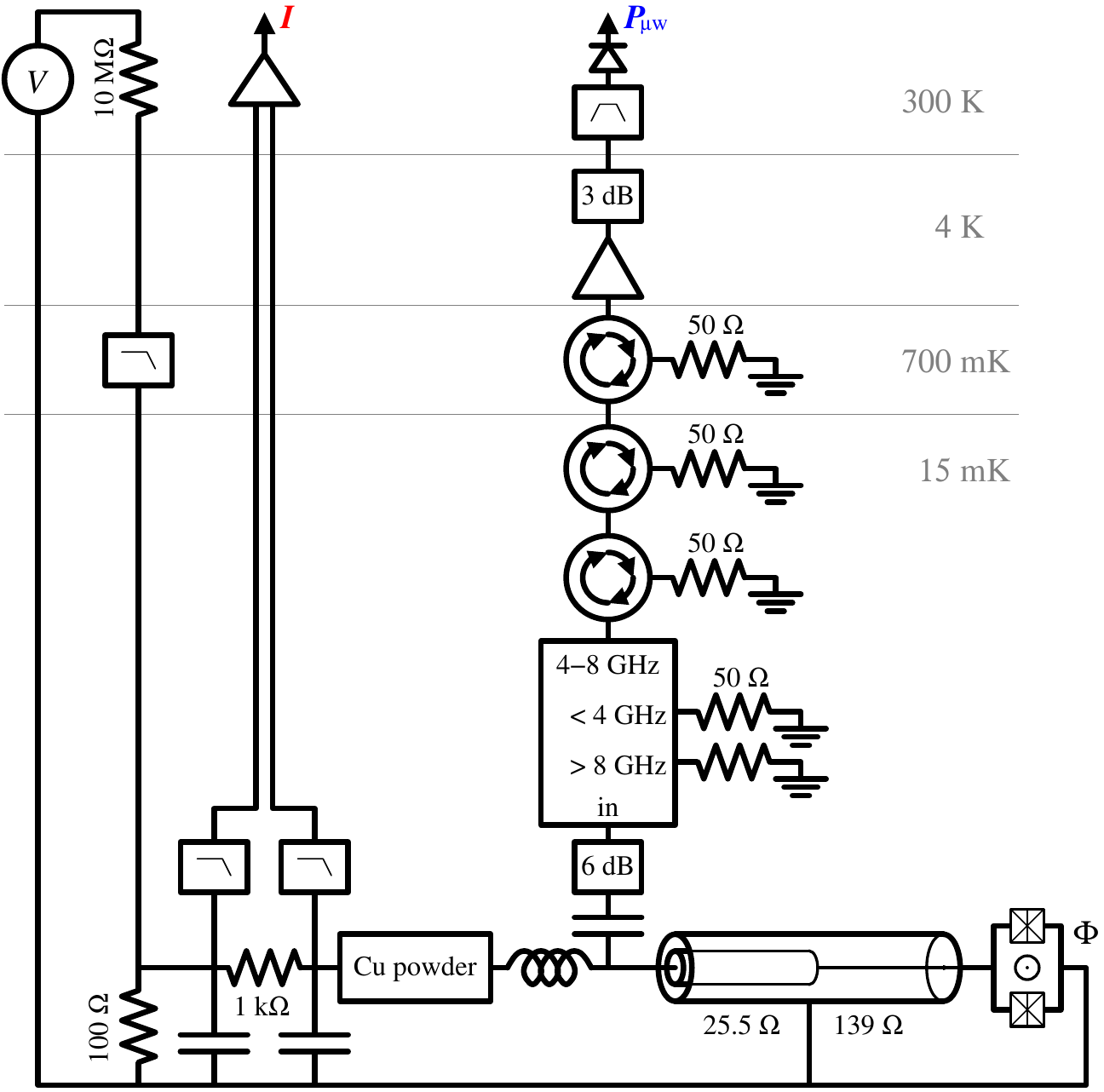} 
\caption{\label{setupdetail} More detailled view of the experimental
  setup. Only circuit components inside the refrigerator are shown in
  full detail.}
\end{figure}

In addition to the components depicted in Fig.~1 in the paper, the low frequency
circuit includes multipole RC low pass filters in the bias line
(cutoff at $\pq{8}{kHz}$) and the current measurement lines (cutoff at
$\simeq\pq{100}{kHz}$), several capacitors to ground and a copper
powder filter between readout resistor and bias T (see Fig.~\ref{setupdetail}).

\emph{Calibration} of the division factor of the voltage divider used for
biasing the sample is performed in situ by reading out the applied
bias voltage through one of the lines used for current measurements
(see Fig.~1). We use a lock-in technique for this measurement and
minimize the critical current of the junction so that the current
through the sample is negligible.

The $\pq{1}{k\Omega}$ NiCr resistor used for current measurement
cannot be calibrated in situ due to large lead resistances, but we
have verified that a resistor from the same batch does not deviate
from its nominal value by more than $\pq{1}{\%}$ at the operating
temperature of our refrigerator.

In the presented measurements we have corrected the applied bias
voltage for the voltage drop over the $\pq{1}{k\Omega}$ resistor.

The good agreement between theory and data in Fig.~4 supports the
accuracy of this DC calibration.

\subsubsection{Microwave circuit}
The microwave chain (see Fig.~\ref{setupdetail}) consists of the
junction, the resonator, a bias T, a $\pq{6}{dB}$ attenuator, a
triplexer with crossover frequencies 4 and $\pq{8}{GHz}$, two 4 to
$\pq{8}{GHz}$ isolators at base temperature, an isolator at
$\pq{700}{mK}$, a cryogenic amplifier at $\pq{4}{K}$ with $\pq{3.5}{K}$
noise temperature, a $\pq{3}{dB}$ attenuator at $\pq{4}{K}$ followed
by a room temperature amplification and measurement chain.  The isolators protect the
sample from the amplifier noise in their working range from 4 to
$\pq{8}{GHz}$. The triplexer, working from DC to $\pq{18}{GHz}$,
connects the sample to the microwave chain only in the 4 to
$\pq{8}{GHz}$ range. Outside this range the sample is connected to
cold $\pq{50}{\Omega}$ loads so that the sample ``sees'' a cold
$\pq{50}{\Omega}$ load throughout the $\pq{18}{GHz}$ frequency range
of the triplexer. Reflections due to the insertion loss of the
triplexer are suppressed by the $\pq{6}{dB}$ attenuator.

\emph{Calibration} of the resonator impedance $Z(\nu)$ and the gain of the microwave chain
from the sample to the top of the refrigerator is performed
in situ because the superconducting resonator must be cooled below its
critical temperature and the amplifier gain and cable attenuation
depend on temperature. We perform both calibrations using
quasiparticle shot noise as white current noise reference. We bias the junction at
$\pq{2}{mV}$, far above the gap voltage $\pq{0.4}{mV}$ of our
junction, where the current noise is in good approximation $S_{II} = e
I$ at frequencies $|\nu| \ll eV/h \simeq \pq{0.5}{THz}$. In order to
separate this noise from the noise floor of the cryogenic amplifier,
we then apply small variations of the bias voltage and measure the
corresponding changes in the measured microwave power with a lock-in
amplifier.

The conversion of $S_{II}$ into emitted microwave power depends on the
environment impedance $Z(\nu)$ seen by the junction and its tunneling
resistance $\Rn$.  First, only a fraction $\Rn^2/|\Rn + Z(\nu)|^2$ of
the current noise is absorbed by the environment, but this factor is
close to 1 ($>0.85$) as $\Rn = \pq{17.9}{k\Omega}$ is much larger than
the environment impedance. The current noise in the environment has
then to be multiplied by $2 \Re Z(\nu)$ to obtain the microwave power
emitted by quasiparticle shot noise:
\begin{equation}
S_P(\nu) = 2eV \frac{\Re Z(\nu) \Rn}{|Z(\nu) + \Rn|^2}.
\label{eq:shotnoise}
\end{equation}
The emitted power depends on $Z(\nu)$, therefore $Z(\nu)$ and the gain
of the microwave chain cannot be calibrated independently. We solve
this problem by making the following assumptions:
\begin{enumerate}
\item The gain of the microwave chain is frequency independent in the
  5 to $\pq{7}{GHz}$ range.
\item The integral over the peak in $\Re Z(\nu)$ from 5 to
  $\pq{7}{GHz}$ corresponds to calculations based on the resonator
  geometry.
\end{enumerate}
Assumption 1 is justified by a previous calibration of the gain of the
amplifiers, found to be constant in the 5 to $\pq{7}{GHz}$ range
within 0.5 dB, and by observing that the noise floor of the amplifier
in this experiment in the same 5 to $\pq{7}{GHz}$ range deviates by
less than $\pq{0.5}{dB}$ from its mean value, meaning that the gain
and attenuation after the input of the cold amplifier are almost
frequency independent. However, frequency dependent attenuation
between the sample and the cryogenic amplifier is not detected this
way and can be source of slight inconsistencies.

Assumption 2 relates to the fact that the integral over $\Re Z(\nu)$,
proportional to the characteristic impedance of the resonance, mainly
depends on the transmission line impedance of the quarterwave segment
closest to the junction, less on the second segment, and only very
weakly on the load impedance beyond the resonator. For example, the
characteristic impedance of the resonance changes by less than
$\pq{20}{\%}$ when the the load impedance is changed from
$\pq{50}{\Omega}$ to any value between $\pq{20}{\Omega}$ and infinity.
Therefore we can get an accurate estimate of this integral from the
impedance of the resonator segments, even in the presence of
imperfections in the line impedance seen from the resonator.  We
calculate the characteristic impedance of our quarterwave segments
from the designed resonator geometry, checked with an electron
microscope and the well known stackup of our sample. We obtain
characteristic impedances $\pq{139}{\Omega}$ and $\pq{25.5}{\Omega}$,
resulting in a characteristic impedance of the resonance mode of
$\pq{150}{\Omega}$.  The fact that $P(E)$ theory using this resonator
impedance accurately fits the relative heights of first and second
order peaks in Fig.~2 indicates that this estimation of the resonator
impedance is correct.

With these two assumptions we can use Eq.~(\ref{eq:shotnoise}) to
separately calibrate the power gain and $Z(\nu)$.  To calibrate the
power gain in Figs.~2 and 3 we compare the measured shot noise signal
to Eq.~(\ref{eq:shotnoise}) integrated over the bandwidth of our
measurement ranging from 5 to $\pq{7}{GHz}$.

In order to determine the spectral dependence of $Z(\nu)$ we use a
narrow band pass filter (as for the data in Fig.~4) to measure a
signal proportional to $S_P$. To extract $\Re Z(\nu)$ from the emitted
power using Eq.~(\ref{eq:shotnoise}), one would need to know $\Im
Z(\nu)$, which is not measured independently. To circumvent this
difficulty, we use Eq.~(\ref{eq:shotnoise}) iteratively to
calculate $Z(\nu)$: we start with the calculated impedance $\Re
Z_0(\nu)$ in the denominator and in each step calculate a new $\Re
Z_n(\nu)$, normalize it using assumption 2, and then calculate $\Im
Z_n(\nu)$ using the Kramers-Kronig relations with $\Re Z_n(\nu)$ in
the 5 to $\pq{7}{GHz}$ range where $S_P$ was measured and $\Re
Z_0(\nu)$ outside this range. This procedure converges quickly and
yields a precise determination of $\Re Z(\nu)$ because $\Im
Z(\nu) \ll \Rn$ and therefore only has a small influence on the extracted $\Re Z(\nu)$.

\subsection{Noise emitted by a Josephson junction in the DCB regime}
We give here two derivations of expression (5) in the paper. The first
one follows the derivation of Coulomb blockade in \cite{ingold92} and
treats photon emission as fluctuations in the Cooper pair tunneling rate. The second derivation explicitly evaluates photon emission and
absorption rates in a mode of the electromagnetic environment.

\subsubsection{Derivation 1}

The Josephson element is described by the Hamiltonian $\Hj = - \Ej
\cos \hat{\delta}$ and the current operator through the element is
\[ \hat{I} = - \frac{2 e}{\hbar}  \frac{\partial \Hj}{\partial \delta} = i
   \frac{e}{\hbar} \Ej \left(e^{i\hat \delta} - e^{-i \hat \delta} \right) \]
The current-current correlator function is then
\begin{eqnarray}
  S_{II} (t) & \!\!=\!\! &\left\langle \hat I(t) \hat I(0)
  \right\rangle \nonumber \\ 
  &\!\! =\!\! & \frac{e^2 E_J^2}{\hbar^2} \left( \left\langle e^{i
    \hat\delta(t)} e^{- i \hat\delta(0)} \right\rangle +
  \left\langle e^{-i\hat\delta(t)} e^{i \hat\delta(0)}
  \right\rangle \right)\!.
     \label{eq:curcor}
\end{eqnarray}
To arrive at the last line we have noted that at non-zero dc bias
voltage $V$ terms $e^{\pm i \hat{\delta}(t)} e^{\pm i
  \hat{\delta}(0)}$ with the same sign in the exponential average to
zero. We decompose the phase difference across the tunnel element
$\hat{\delta} (t) = \delta_0 ( t) + \tilde{\delta} ( t)$ into the
deterministic part $\delta_0 (t) = 2 eVt / \hbar$ and a fluctuating
random phase $\tilde{\delta} ( t)$ caused by the fluctuations in the
electromagnetic environment. We use the gaussian
character of the noise of the linear environment to rewrite \cite{ingold92}:
\begin{eqnarray*}
  \left\langle e^{\pm i \hat{\delta} ( t)} e^{\mp i \hat{\delta} ( 0)}
  \right\rangle &=& \text{$\text{$e^{\pm i 2 e V t / \hbar}$} \left\langle
  e^{\pm i \tilde{\delta} ( t)} e^{\mp i \tilde{\delta} ( 0)} \right\rangle$}\\
  & = & \text{$e^{\pm i 2 e V t / \hbar}$} e^{J ( t)}
\end{eqnarray*}
where $J ( t) = \left\langle \left( \tilde{\delta} ( t) - \tilde{\delta} ( 0)
\right)  \tilde{\delta} ( 0) \right\rangle$ is the (superconducting)
phase-phase correlation function across the impedance, so that
(\ref{eq:curcor}) becomes
\[ S_{II} ( t) = \frac{e^2 E_J^2}{\hbar^2} 2 \cos \frac{2 eVt}{\hbar} e^{J (
   t)} . \]
The (non-symmetrized) current noise density $S_{II}(\nu)$ is
directly the Fourier transform of the current--current correlator $S_{II} ( t)$
\[
S_{II}(\nu) 
= \int_{-\infty}^\infty S_{II}(t) e^{-i 2 \pi \nu t} dt \label{eq:SII} 
\]
(Wiener Khinchin theorem) which yields immediately
\begin{eqnarray*}
  S_{II} ( \nu, V) & = & \frac{2 \pi e^2 E_J^2}{\hbar^{}} \left(
  P(2eV-h\nu) + P (-2eV-h\nu)\vphantom{\tilde{0}}
  \right) \label{eq:noise}
\end{eqnarray*}
with $P ( E) = \frac{1}{2 \pi \hbar} \int_{- \infty}^{\infty} e^{J (
  t) + iEt / \hbar} dt$ \ the Fourier transform of $\exp J(t)$.
Assuming that the mode of the environment at frequency $\nu$ is in the
ground state, i.e.\ $kT \ll h\nu$, the current noise gives rise to a
photon emission rate density
\begin{eqnarray*}
\density &=& \frac{2\Re Z(\nu)}{h\nu} S_{II}\\ 
& =& \frac{2\znorm(\nu)}{\nu}
\frac{\pi}{2\hbar}\Ej^2 
\left(P(2eV-h\nu)+P(-2eV-h\nu)\vphantom{\tilde{0}}\right).
\end{eqnarray*}

In expression~(5) in the paper we have neglected the second term
corresponding to photon emission during backward tunneling which is
negligible for $kT \ll 2eV + h\nu$ as in our experiment.

\subsubsection{Derivation 2}
The Hamiltonian of the total system is 
\begin{equation*}
\hat H = \hat H_{\rm em} + \Hj
\end{equation*}
with
\begin{eqnarray*}
\hat H_{\rm em} &=& \sum_l \frac{\left(\frac{\hbar}{2e}\hat \phi_l\right)^2}{2L_l} + \frac{\hat Q_l^2}{2C_l}\\
\Hj &=& -E_{\rm J} \cos \hat \delta\\
\hat \delta &=& \frac{2eVt}{\hbar} + \sum_l \hat \phi_l.
\end{eqnarray*}
Here $\hat H_{\rm em}$ is the Hamiltonian describing the environment
of impedance $Z(\omega)$. $C_l$ and $L_l$ describe the mode at
frequency $\nu_l = 1/2\pi \sqrt{C_l L_l}$. A second constraint on $C_l$ and $L_l$ imposed by the requirement that the sum of the resonator
impedances approaches $Z(\nu)$, giving
\begin{equation*}
Z_l = \sqrt{\frac{L_l}{C_l}} = \frac{2}{\pi} Z(\nu) \frac{\Delta \nu}{\nu},
\end{equation*}
where $\Delta \nu$ is the spacing between adjacent modes, supposed to be constant.
The Cooper pair tunneling rate can now be calculated using Fermi's golden rule:
\begin{eqnarray*}
\Gamma^\rightarrow& = &\frac{\pi}{2\hbar} E_{\rm J}^2 \!\!\sum_{\begin{smallmatrix}n_l &\!\ge\!& 0\\ \!\!\!\!\!\!\!\!\!\!\!\!n_l + m_l &\!\ge\!& 0\end{smallmatrix}} \!\!\prod_{l} \occup_{l,n_l}
\left|\bracket{n_l+m_l}{e^{i\hat
    \phi_l}}{n_l}\right|^2\\ &&\hspace{\fill} \delta\!\left(\!\sum_l
m_l h \nu_l - 2eV\!\!\right)\\&=:& \frac{\pi}{2\hbar} \Ej^2 P(2eV),
\end{eqnarray*}
where $\occup_{l,n}$ is the probability for mode $l$ to be in state
$\ket{n}$ before the tunneling event. We assume that the tunneling
rate is low so that the environment can reach thermal equilibrium
before each tunneling event and the $\occup_{l,n}$ describe thermal
equilibrium states. This equation can be seen as the definition of
$P(E)$, the probability density for the environment to absorb energy
$E$ when a Cooper pair tunnels. Calculating $P(E)$ by evaluating this
sum is possible but cumbersome, so usually the approach from
Ref.~\cite{ingold92} is taken.

In order to calculate the rate for emitting photons at frequency $\nu$
we treat mode $k$ at $\nu_k = \nu$ separately and keep track of
the photon number difference $m_k$. We absorb all other modes into a
function $P'(E)$ describing a modified environment where mode $k$ has
been removed.
\begin{equation*}
\Gamma^\rightarrow_{m} = \!\!\!\!\!\!\!\!\!\!\sum_{n=\max\{0,-m\}}^\infty\!\!\!\!\!\!\!\!\!\!\occup_{k,n}\left|\bracket{n\!+\!m}{e^{i\hat
    \phi_k}}{n}\right|^2 \frac{\pi}{2\hbar} E_{\rm J}^2 P'(2eV\!-\!m h
\nu_k).
\end{equation*}
In this expression the phase operator $\hat \phi_k$ across the
resonator $k$ can be expressed in terms of the rising and lowering
operators $a^\dagger_k$ and $a_k$:
\begin{equation*}
\hat \phi_k = \sqrt{\prefactor} (\hat a^\dagger_k + \hat a_k)
\end{equation*}
with
\begin{equation*}
\prefactor = \pi \frac{4e^2}{h} Z_k = 2 \frac{4e^2}{h} \Re Z(\nu)
\frac{\Delta \nu}{\nu} = 2 \znorm(\nu) \frac{\Delta \nu}{\nu}
\end{equation*}
The spectral density of photon emission/absorption is calculated in
the limit $\Delta \nu \rightarrow 0$ where $e^{i\hat \phi_k}
\rightarrow 1 + i\sqrt{\prefactor}\left(\hat a^\dagger_k + \hat
a_k\right)$ and $P'(E) \rightarrow P(E)$. In this limit
$\Gamma^\rightarrow_m$ is nonzero only for $m = 0, \pm 1$. The rate
without photon emission or absorption $\Gamma^\rightarrow_0$ simply
tends to the bare Cooper pair tunneling rate,
i.e.\ $\Gamma^\rightarrow_0 \rightarrow \Gamma^\rightarrow$. For
$m=\pm1$ we calculate the density of photon absorption/emission
\begin{eqnarray}
\density^\rightarrow_\pm(\nu) &=& \lim_{\Delta \nu \rightarrow 0}
\frac{\Gamma^\rightarrow_{\pm 1}}{\Delta \nu} \\ &=& 2
\frac{\znorm(\nu)}{\nu} \sum_{n=1}^\infty n \occup_{n^{-1}_{+0}}(\nu)
\frac{\pi}{2\hbar} E_{\rm J}^2 P(2eV \mp h \nu).\nonumber
\end{eqnarray} 
The backward tunneling processes $\Gamma^\leftarrow$ and
$\density^\leftarrow_{\pm 1}$ have the same expressions as the
corresponding forward processes but with $V$ replaced by $-V$.

At low temperature $kT \ll h\nu$ the mode at $\nu$ is in the thermal
ground state, i.e.\ $\occup_n (\nu) \simeq \delta_{n,0}$, and
$P(-2eV-h\nu) \simeq 0$, so that $\density^\rightarrow_- =
\density^\leftarrow_\pm = 0$. We then find expression (5) in the
paper:
\begin{equation*}
\density(\nu) = \density^\rightarrow_+(\nu) = 
 \frac{2\znorm(\nu)}{\nu}  \frac{\pi}{2\hbar} E_{\rm J}^2 P(2eV - h \nu)
\end{equation*}

\end{document}